\documentclass[12pt]{article}
\usepackage[english]{babel}
\usepackage{amssymb}
\usepackage{amsmath}
\usepackage{graphicx}
\usepackage{pazha}
\usepackage{natbib}
\tightenlines

\sloppypar

\begin{document}
\baselineskip 21pt

\title{\bf Azimuthal patterns in planetesimal circumstellar disks}

\author{\bf \hspace{-1.3cm} \ \
Tatiana\,Demidova$^{1,2}$\affilmark{*}, Ivan\,Shevchenko$^{2,3}$}

\affil{
{\it $^{1}$ Crimean Astrophysical Observatory RAS, Nauchniy \\
$^{2}$ St. Petersburg State University,
Universitetskaya emb. 7/9, 199034 St. Petersburg \\
$^{3}$ Institute of Applied Astronomy RAS, nab. Kutuzova 10,
191187 St. Petersburg}}

\vspace{2mm}
%\received{~~~~~~~~}
%\received{\today}
%\revised{\today}

\sloppypar \vspace{2mm}

{\bf Abstract.} Ways of formation of azimuthal resonant patterns
in circumstellar planetesimal disks with planets are considered.
Our analytical estimates and massive numerical experiments show
that the disk particles that initially reside in zones of
low-order mean-motion resonances with the planet may eventually
concentrate into potentially observable azimuthal patterns. The
structuring process is rapid, usually taking $\sim$100 orbital
periods of the planet. It is found that the relative number of
particles that retain their resonant position increases with
decreasing the mass parameter $\mu$ (the ratio of masses of the
perturbing planet and the parent star), but a significant fraction
of the particle population is always removed from the disk due to
accretion of the particles onto the star and planet, as well as
due to their transition to highly elongated and hyperbolic orbits.
Expected radio images of azimuthally structured disks are
constructed. In the considered models, azimuthal patterns
associated with the 2:1 and 3:2 resonances are most clearly
manifested; observational manifestations of the 1:2 and 2:3
resonances are also possible.


%\vfill
\noindent\rule{8cm}{1pt}\\
{$^*$e-mail: $<$proxima1@list.ru$>$}

\newpage

%\vspace{1cm}

\section*{Introduction}

Presence of planets in a debris planetesimal disk has a
significant effect on the low-mass matter distribution inside the
disk. Mean-motion resonances with a planet form ring-shaped dense
structures as well as matter-free cavities in the disk
\citep{2000ApJ...537L.147O, 2002ApJ...578L.149Q,
2003ApJ...588.1110K, 2006MNRAS.373.1245Q, 2012MNRAS.419.3074M,
2015ApJ...799...41M, 2016MNRAS.463L..22D}. Planetary perturbations
can form the disk boundaries, both external and internal,
depending on the system configuration \citep{1999ApJ...527..918W,
2006MNRAS.372L..14Q, 2013ApJ...763..118S, 2014ApJ...780...65R}.

The planetesimal (debris) circumstellar disk is formed together
with the planetary system of a star as a result of the evolution
of the protoplanetary gas-dust disk \citep[see, e.g., the review
by][]{2022arXiv220203053M}. The debris disk consists of solid
bodies with sizes in a broad range, from fine micron-sized dust to
kilometer-sized and larger planetesimals.

Previously, we considered the formation of spiral and ring
patterns in debris disks
\citep{2015ApJ...805...38D,2016MNRAS.463L..22D}. Here we study an
{\it azimuthal} structuring by the planet of the debris disk
matter that occur in the vicinities of resonances of mean motions
(mean orbital frequencies) of minor bodies with the planet
\citep{2000ssd..book.....M, 2020ASSL..463.....S}. We consider
first-order resonances, namely, the 2:1, 3:2, 1:2, and 2:3 ones.

\section*{Formation of resonant azimuthal patterns}

\begin{figure}
\centering \includegraphics[width=0.8\textwidth]{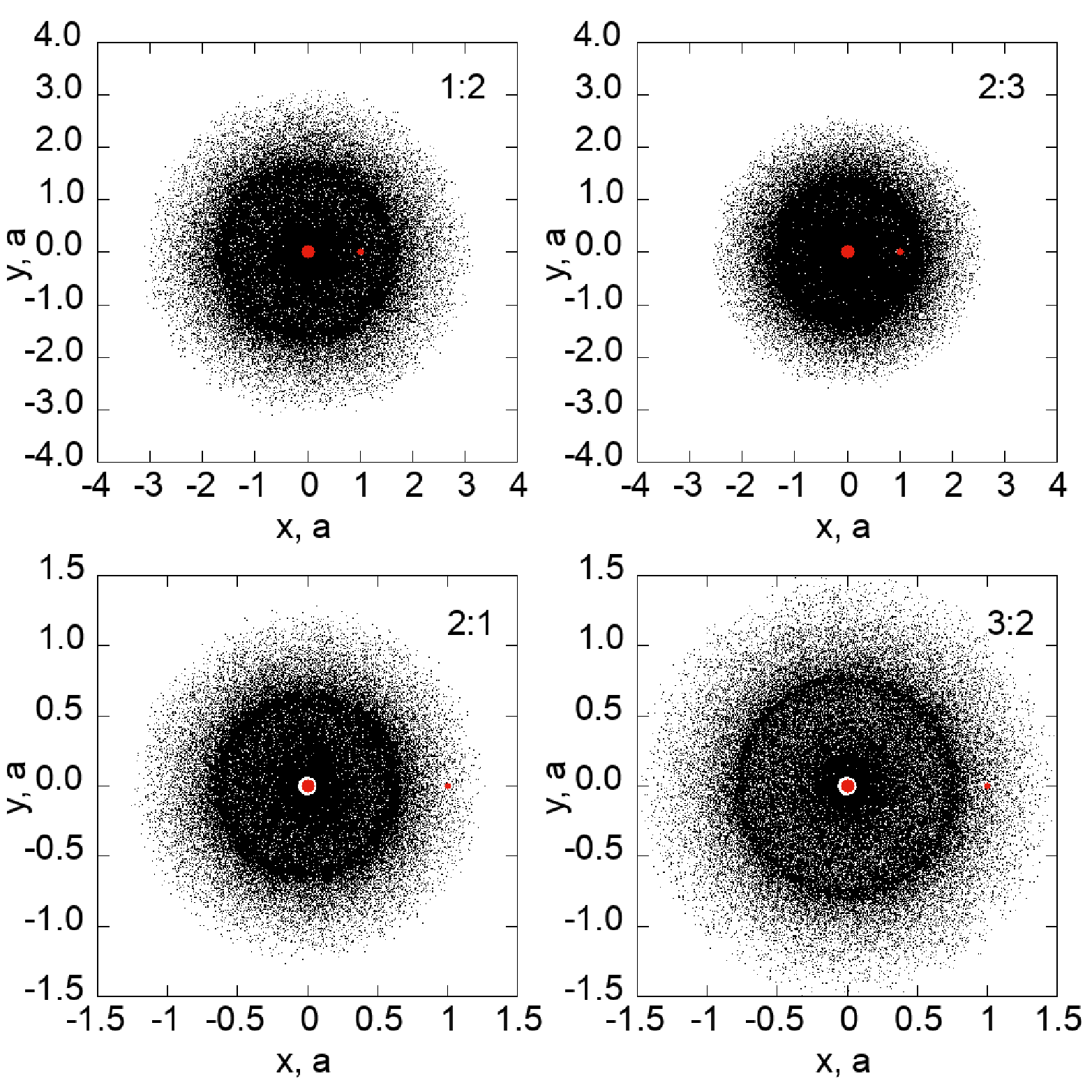}
\caption{The initial ($t=0$) distribution of particles at
mean-motion resonances 1:2, 2:3, 2:1, and 3:2, in a model with the
mass parameter $\lg\mu=-3$. The locations of the star and planet
are marked with red dots. The coordinates along the axes are given
in units of the semimajor axis of the planet's orbit.}
\label{fig:particles0}
\end{figure}

Note that, according to \citet{2014MNRAS.442.1755K}, the
circumstellar disk surface density distribution, which consists of
many Keplerian orbits, may have maxima near the inner and outer
boundaries of the disk. The density maximum at the inner boundary
is due to the crowding of orbits near the pericenters. A similar
density thickening can be seen in the initial distribution of
particles in Fig.~\ref{fig:particles0}. Emergence of the maximum
at the apocenter is due to the particle dynamics, and therefore
does not manifest itself in the particle initial distribution.

It is known that the velocity of a body in an elliptical orbit is
maximum at the orbit pericenter and minimum at the apocenter. The
velocity values are, respectively, given by
\begin{equation}
  v_\mathrm{p} = na \sqrt{\frac{1 + e}{1 - e}} , \qquad
  v_\mathrm{a} = na \sqrt{\frac{1 - e}{1 + e}} ,
\end{equation}

\noindent where $n$, $a$, and $e$ are the orbital mean motion,
semimajor axis, and eccentricity, respectively; see, e.~g.,
formulas (2.35) in \cite{2000ssd..book.....M}. Therefore, the
ratio of the maximum and minimum velocities is
\begin{equation}
  v_\mathrm{p}/v_\mathrm{a} = \frac{1 + e}{1 - e} ,
\end{equation}

\noindent and, accordingly, the ratio of the particle residence
times in the vicinities of the apocenter and pericenter is
\begin{equation}
  T_\mathrm{a}/T_\mathrm{p} \approx \frac{1 + e}{1 - e} , \label{eqT}
\end{equation}

\noindent which tends to infinity at $e \to 1$.

Planetary perturbations affect the particle surface density
distribution. Our calculations show that the area at the disk's
inner boundary is cleared from matter.

Therefore, an external observer of an ensemble of particles in
elongated orbits would see the particles mostly concentrated
towards their orbital apocenters. This simple fact, as shown
below, substantiates the opportunity for observational
manifestation of azimuthal patterns in the disk.

In what follows, we assume that any mutual perturbations of
particles in orbits are absent.

We consider the trajectories of particles in various resonances in
a rotating (with the angular velocity of the perturbing planet)
coordinate system. The planet's orbit is assumed to be circular.
Let us consider internal and external resonances of the first
order.

If the particle resides in an internal $p+q:p$ resonance with the
planet, then the ``star--planet--particle'' configuration is
repeated on the time intervals equal to $p+q$ particle's orbital
periods; if the particle resides in an external $p:p+q$ resonance,
then the configuration is repeated on the time intervals equal to
$p$ particle's orbital periods \citep{2000ssd..book.....M}.

In \cite{2000ssd..book.....M}, particle trajectories are shown for
some first-order resonances with the planet; see Fig.~8.4 in that
book. In a rotating (with the planet's angular velocity)
coordinate system, the particle trajectories possess specific
features, namely, loops, which appear either near the apocenter
(for inner particles) or near the pericenter (for outer
particles).

For the ``star--planet--particle'' configuration be once again
repeated, at the $p+q:p$ internal resonance the particle should
accomplish $p+q$ orbital revolutions, and, at the $p:p+q$ external
resonance, $p$ revolutions; therefore, the number of loops of its
trajectory would be, respectively, $p+q$ and $p$.

The angular velocity of a planet in a circular orbit is constant,
while that of a particle in an elliptical orbit varies with time.
If, starting from zero, one increases the eccentricity of an inner
particle's orbit, then, at some critical eccentricity value, the
particle's angular velocity at the apocenter will coincide with
the constant angular velocity of the planet, and, in the rotating
coordinate system, the particle trajectory  will acquire a cusp.
If the eccentricity is increased further on, then the particle
trajectory forms a backward loop.

It is straightforward to show \citep[see][]{2000ssd..book.....M}
that for an inner particle in the $p+q:p$ resonance with the
planet the cusp appears when its eccentricity $e$ satisfies the
cubic equation
\begin{equation}
(1 + e)^3 = [(p + q)/p]^2(1 - e) , \label{eq1}
\end{equation}
and, for an external particle in the $p:p+q$ resonance with the
planet, it appears at $e^\prime$ satisfying the cubic equation
\begin{equation}
(1 - e^\prime)^3 = [p/(p + q)]^2(1 + e^\prime) . \label{eq2}
\end{equation}
For example, solving the Eq.~(\ref{eq1}) for the 2:1, 3:2, and 4:3
internal resonances gives the critical values of $e$ equal to
0.365, 0.211 and 0.148, respectively \citep{2000ssd..book.....M}.

According to the Eq.~(\ref{eqT}), the maximum contrast in the
observed apocentric and pericentric concentrations of particles
should be expected in the first of these cases (resonance 2:1),
and the minimum contrast is in the third case (resonance 4:3).

In the Solar system, an actual example of a resonant group,
forming a weakly pronounced quasi-triangular rotating (with the
angular velocity of Jupiter) pattern in the main asteroid belt, is
provided by the Hilda group \citep{LSST04}, which resides in the
3:2 resonance with Jupiter.

\section*{The model and methods}

Let us consider the disk dynamics in the planar problem in the
barycentric frame. Let the system include a central star with mass
$M=M_{\odot}$, a planet with mass $m$ in a circular orbit with
period $P$ and semimajor axis $a$, and a disk of passively
gravitating particles, the parameters of which will be specified
below.

Initially, the star is situated at the point with coordinates
($x$, $y$) = ($p_1$, 0), and the planet is at the point ($p_2$,
0), where $p_1=-\frac {m}{M + m} a$ and $p_2=\frac{M}{M + m} a$.
The velocity vectors of the star and the planet initially have
components (0,$-\frac{m}{M+ m}na$) and (0,$\frac{M}{M + m}na$),
respectively, where the mean motion $ n =\frac{2\pi}{P}$. The
model mass parameter $\mu=\frac{m}{M+m}$ is varied: the system
dynamics is considered at its four values: $\lg\mu=-2,-3,-4$ and
$-5$.

Initially, $10^6$ massless (passively gravitating) particles are
placed at a given resonance. In \citep{2020AstL...46..774D},
radial sizes of the ring-like chaotic zones at mean-motion
resonances were estimated in dependence on the mass parameter:
$\Delta a_\mathrm{r}=0.91\mu^{0.43}a$ for internal (with respect
to the planet's orbit) resonances, and $\Delta
a_\mathrm{r}=1.32\mu^{0.42}a$ for external ones. Here we use these
$\Delta a_\mathrm{r}$ values to determine the boundaries for the
initial values of the semimajor axes of particle orbits: the
particles are initially distributed randomly along the semimajor
axis within $a_\mathrm{r}\pm\frac{\Delta a_r} {2}$, where
$a_\mathrm{r}$ is the resonance location. This choice of the
boundaries is called forth by the fact that the particles are
effectively ejected from the chaotic zone (where their
concentration decreases with time, see
\citealt{2020AstL...46..774D}), and thus can participate in
formation of patterns.

There are also cosmogonic reasons for such a choice: if any planet
in the disk were originally formed close to resonance with another
planet and were destroyed by colliding with another object, e.~g.
planetary embryo (on collisional evolution during the formation
stages of planetary systems, see \citet{ 1999Icar..142..219A,
1998Icar..136..304C, 1998M&PSA..33S.153T}), this would contribute
to an increase in the concentration of planetesimals in the
resonance zone.

The calculations were performed for four mean-motion
particle--planet resonances: two outer ones, 1:2 and 2:3, and two
inner ones, 2:1 and 3:2. The initial eccentricities of particles
are distributed randomly in the range $[0:1)$, and the initial
true anomaly and longitude of pericenter values are randomly
distributed in the range $[0,2\pi)$.

The calculations were performed in the planar problem in
rectangular barycentric frames. The planet and particles move in
their orbits counterclockwise. The integration of the particle
motion equations was carried out using the Bulirsch--Stoer
algorithm \citep{1980ina.book.....P, 1992nrca.book.....P}. The
error tolerance $\epsilon$ was set equal to $10^{-14}$. In the
course of calculations, the Jacobi integral conservation was
controlled for each particle. The method implementation is
described in detail in \citet{2022A&C....4100635D}.

The disk particles may escape from the system due to three basic
processes: accretion onto the planet, accretion onto the star, and
scattering to an elongated orbit. The accretion radius for the
star was set as
$R_\mathrm{sa}=0.01\Big(\frac{M}{3m}\Big)^\frac{1}{3}a$ and that
for the planet as
$R_\mathrm{pa}=0.01\Big(\frac{m}{3M}\Big)^\frac{1}{3}a$. Any
particles that approached the star or the planet within a distance
less than the corresponding accretion radius were considered to
have been accreted and were removed from the system. Any particles
that moved away from the barycenter to a distance of more than 4
orbital radii of the planet were also regarded as escaped from the
system. The maximum duration of integration of any particle orbit
was set to $10^5$ orbital periods of the planet.

In this study, the self-gravity of the planetesimal disk was not
taken into account. \citet{2014A&A...561A..43B} and
\citet{2014MNRAS.443.2541P} showed that the influence of a massive
planet dominates over the mutual gravitational interactions of
planetesimals if the planet's mass is an order of magnitude (or
more) greater than the disk's mass. The planet's minimum mass that
we considered in this study is equal to $10^{-5}M_\odot$ ($\sim 3$
Earth masses), and, for the planetesimals, only a ring of matter
close to resonance was considered. Such a planet's mass is an
order of magnitude greater than the mass of the Kuiper belt and
several orders of magnitude greater than the mass of the main
asteroid belt, estimated in
\citep{2018AstL...44..554P,2018CeMDA.130...57P}. On this reason we
assume that disk's self-gravity can be neglected.

\section*{Particle dynamics}

As soon as the distribution of particles and the rates of their
removal from the system at all considered values of $\mu$ are
respectively similar, solely the case of $\lg\mu=-3$, unless
otherwise indicated, will be described in detail in what follows.

\subsection*{Particle distributions}

Fig.~\ref{fig:particles0} shows the initial distribution of
particles in each model. Initially, the matter is concentrated in
a ring where the corresponding resonance is located.

\begin{figure}
\centering \includegraphics[width=0.8\textwidth]{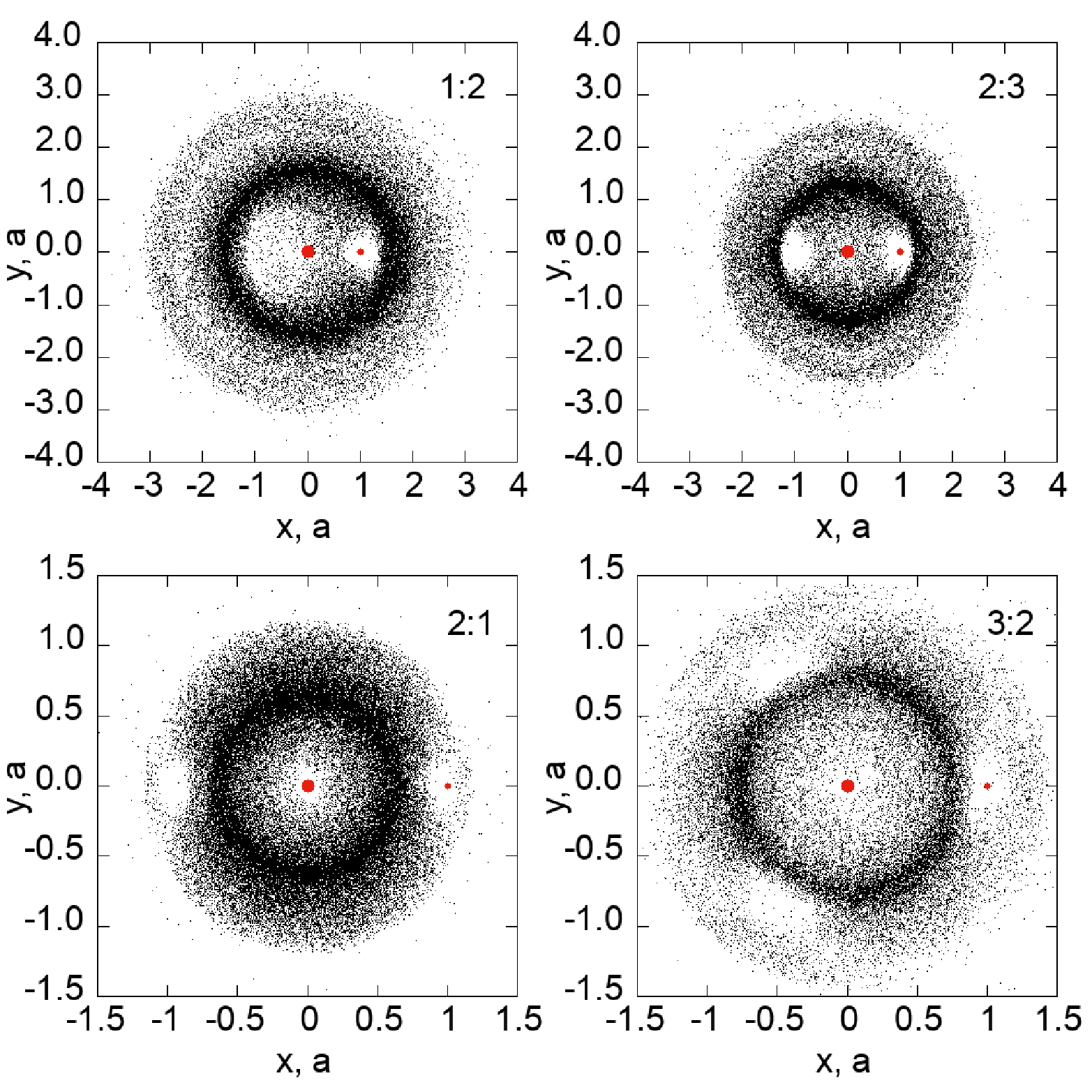}
\caption{Same as Fig.~\ref{fig:particles0}, but at time
$t=10^5P$.} \label{fig:particles}
\end{figure}

Due to planetary perturbations, the particle distribution acquires
azimuthal patterns. Fig.~\ref{fig:particles} shows the particle
distribution at the end of simulation, at time $t=10^5P$ (where
$P$ is the planet's orbital period). At the 1:2 resonance
(Fig.~\ref{fig:particles}, upper-left), a dense ring near the
resonance location at $R\approx 1.58a$ (where $a$ is the planet's
orbital radius) is preserved, but this ring-like pattern is
already asymmetric: the particles condense behind the planet in
the $x$ axis, in its positive part. An area inside the ring
becomes rarefied; asymmetric zones of reduced particle density are
noticeable near the star and the planet. In addition, an arc-like
region of reduced density emerges near $R=2.4a$ (i.~e., near the
2:7 resonance), located symmetrically with respect to the negative
part of the $x$ axis; the outer edge of the disk is asymmetric.

If the particles are initially distributed near the 2:3 resonance
(Fig.~\ref{fig:particles}, upper-right), the resulting annular
pattern has two symmetric clusters along the $y$ axis. Besides,
near the $y$ axis, two arc-like regions with reduced density
emerge at $R\approx 2.08a$, corresponding to the 1:3 resonance. In
the inner (with respect to the resonance location) zone, the
matter is rarefied, and two symmetric cavities emerge, one near
the planet and one on the opposite part of the $x$ axis.

Near the 2:1 resonance (Fig.~\ref{fig:particles}, lower-right),
the symmetric ring is preserved. The matter in close-to-star space
is rarefied. Besides, two cavities emerge, one near the planet,
and a symmetric one on the opposite part of the $x$ axis, as we
have already seen in the 2:3 resonance case.

The most interesting azimuthal pattern emerges if the matter is
initially distributed near the 3:2 resonance. The pattern has
three matter condensations; and cavities are also formed: one of
them surrounds the planet, and the other two are azimuthally
displaced with respect to the planet by $\pm120^\circ$. The matter
in close-to-star space is also rarefied, as observed already in
the previous cases.

Our computations show that the disk becomes structured on the time
scale $\sim 100P$, and afterwards the picture changes with time
only marginally.

\subsection*{Escaping particles}

The time dependences of the number of escaping (leaving the
system) particles manifest themselves similarly in all three basic
processes of escape: in the beginning, a sharp increase in the
number of escaping particles is observed, and afterwards the
dependence eventually reaches a horizontal plateau. In the
processes of accretion onto the planet and scattering to distant
orbits, the plateau is reached at time $\sim 400P$; whereas in the
processes of accretion onto the star and transition to hyperbolic
orbits, at $\sim 200P$. Thus, in the efficiency of particle
removal, the first two processes substantially dominate.

\begin{figure}
\centering \includegraphics[width=0.8\textwidth]{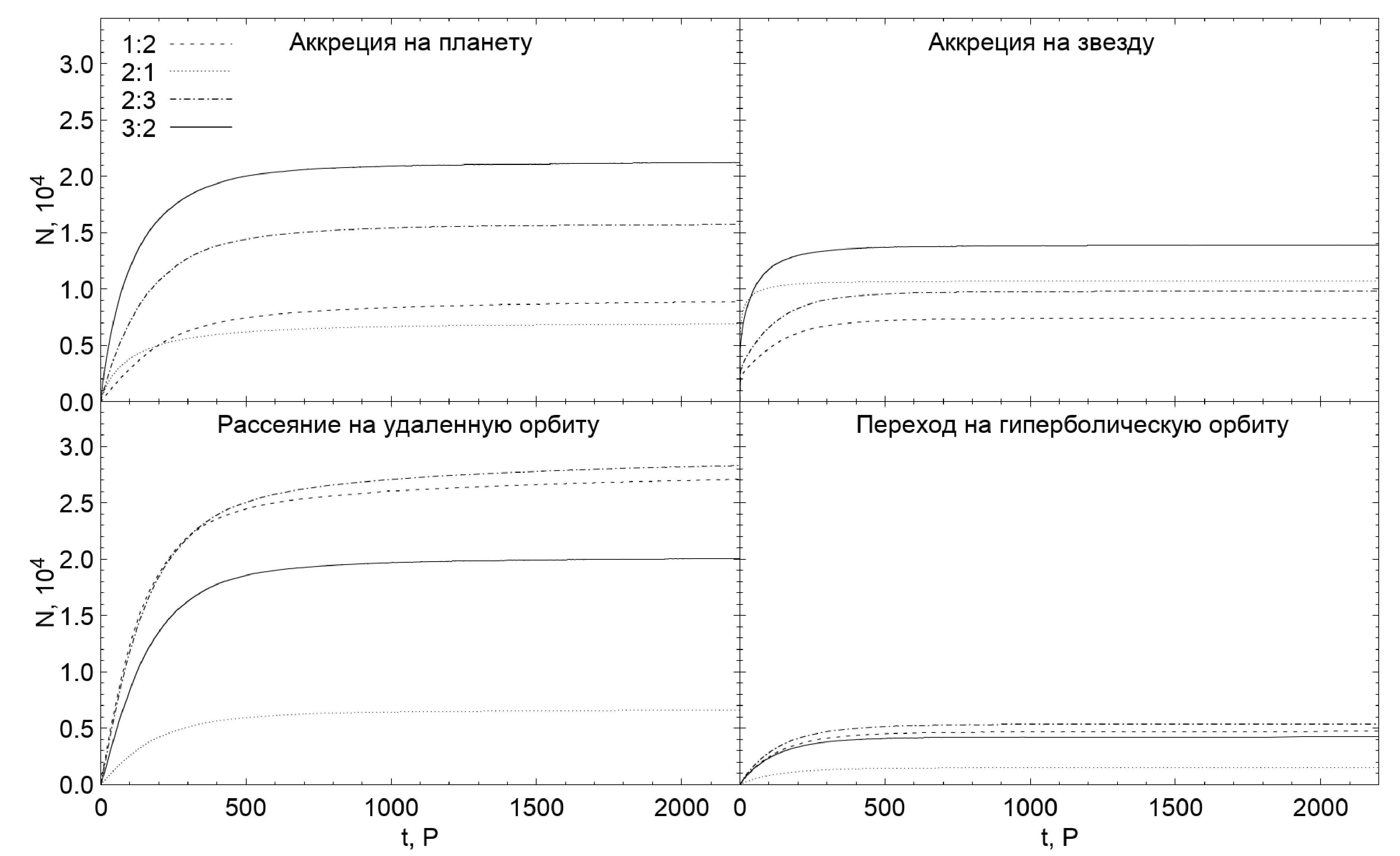}
\caption{Number $N$ of particles escaping (leaving the system)
over time $t$: due to accretion onto the planet (upper-left), due
to accretion onto the star (upper-right), scattering to distant
orbits (lower-left), transition to hyperbolic orbits
(lower-right); the mass parameter $\lg\mu=-3$ in all cases. The
curves correspond to the dependences for particles initially
distributed in zones of resonances 1:2 (dashed), 2:1 (dotted), 2:3
(dash-dotted), and 3:2 (solid). The number of particles is given
in $10^4$ units, and time is in the planet's orbital periods.}
\label{fig:rem}
\end{figure}

In what concerns accretion onto the planet, mostly particles from
the 3:2 and 2:3 resonance zones accrete, since these resonances
are close to the planet's orbit; particles from the internal
resonance 3:2 are falling onto the planet much more often than
those from the external resonance 2:3. Particles that are
initially located near the 1:2 and 2:1 resonances provide
approximately similar contributions to the accretion onto the
planet (Fig.~\ref{fig:rem}, upper-left).

In the process of accretion onto the star, particles from internal
resonances dominate, but for the 2:1 and 2:3 resonances the
difference is minor. Here, as in the case of accretion onto the
planet, the 3:2 resonance particles dominate (Fig.~\ref{fig:rem},
upper-right).

In the process of scattering to distant orbits, noticeably those
particles mostly contribute that are initially located near
external resonances, as one would expect. The dependences for the
1:2 and 2:3 resonances are very similar. Particles from the 3:2
resonance zone also substantially contribute to this scattering
process (Fig.~\ref{fig:rem}, lower-left).

Approximately $10\%$ of particles that have crossed the $R=4a$
limit are transferred to hyperbolic orbits, and can be therefore
regarded as ejected from the system. Particles from the 1:2, 2:3,
and 3:2 resonance zones contribute almost similarly to this
process, whereas in the 2:1 resonance case the relative number of
particles ejected to hyperbolic orbits is much less; see
Fig.~\ref{fig:rem}, lower-right panel.

\begin{figure}
\centering \includegraphics[width=0.8\textwidth]{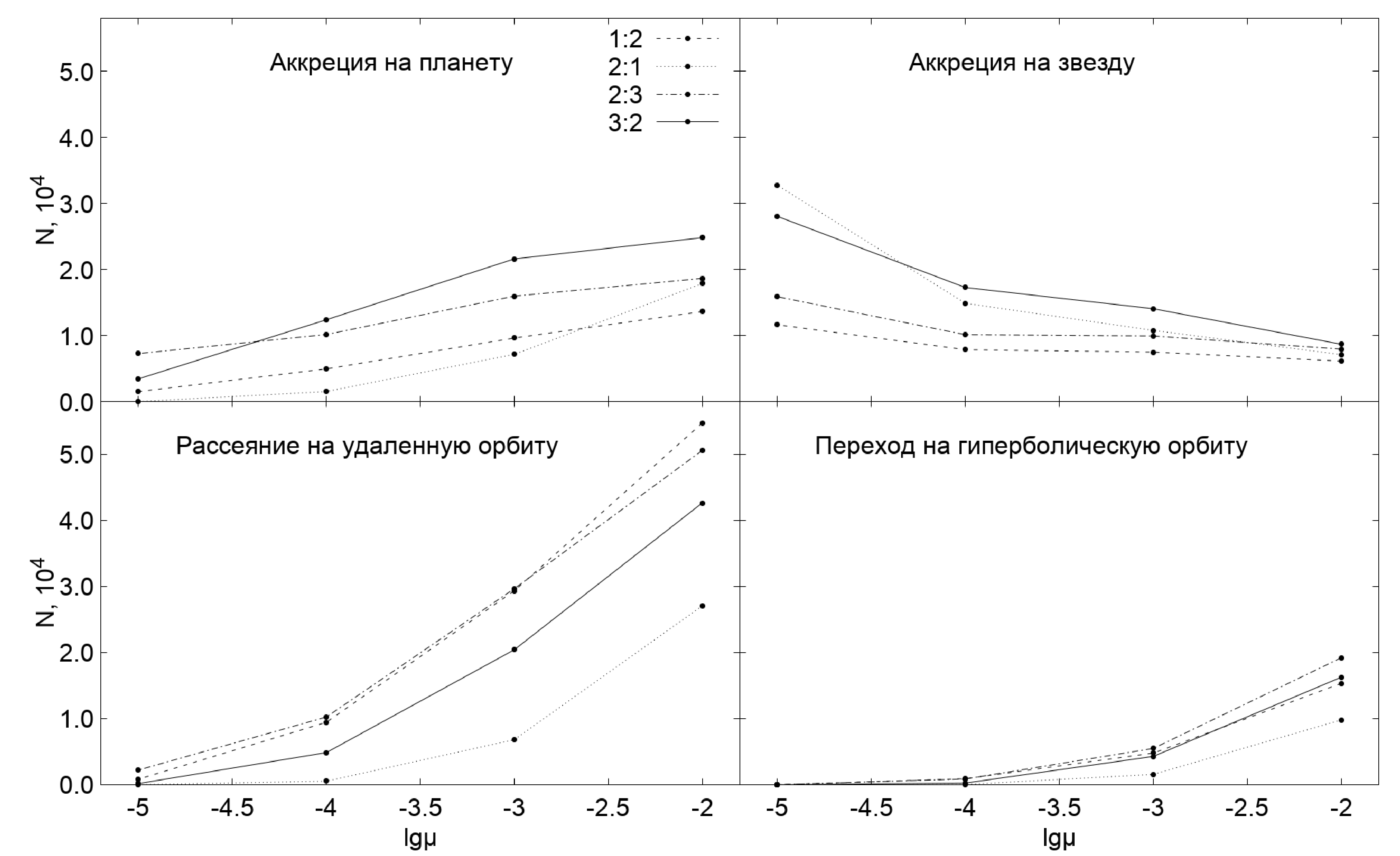}
\caption{Number $N$ of particles escaping over time $t=10^5P$, in
dependence on the mass parameter $\mu$: due to accretion onto the
planet (upper-left), accretion onto the star (upper-right),
scattering to distant orbits (lower-left), transition to
hyperbolic orbits (lower-right). As in the previous Figure, the
curves correspond to the dependences for particles initially
distributed in zones of resonances 1:2 (dashed), 2:1 (dotted), 2:3
(dash-dotted), and 3:2 (solid). The number of particles is given
in $10^4$ units.} \label{fig:remAll}
\end{figure}

Fig.~\ref{fig:remAll} presents the number of particles that have
left the system, as summed at the end of computation, in all our
models. It can be seen that the number of particles accreting onto
the planet, scattering to distant orbits, and passing to
hyperbolic orbits increases with increasing the mass parameter
$\mu$. Conversely, in the case of accretion onto the star, the
observed dependence is opposite; apparently, this is determined by
an increase in the number of particles leaving the computation
domain in other ways. Note that the relative number of particles
that have left the computation domain in each of the processes
coincides with that presented above in Fig.~\ref{fig:rem}. Also
note that for the minimum used here mass parameter value, namely
$\lg\mu=-5$, the particles are not transferred to hyperbolic
orbits in all considered resonance cases.

\section*{Observability of azimuthal patterns}

\begin{figure}
\centering \includegraphics[width=0.8\textwidth]{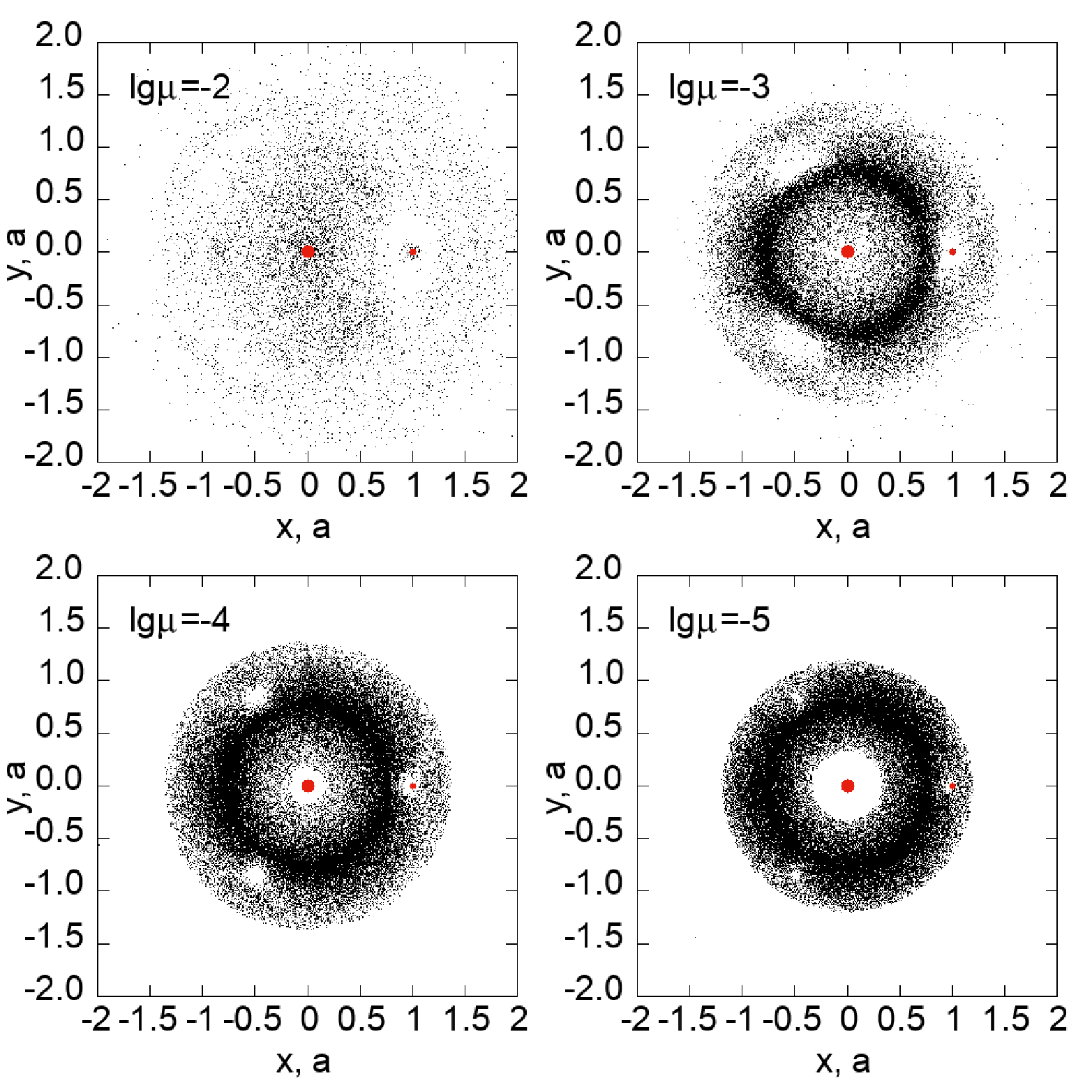}
\caption{Same as Fig.~\ref{fig:particles}, but for resonance 3:2;
in models with the mass parameter $\lg\mu=-2$ (upper-left),
$\lg\mu=-3$ (upper-right), $\lg\mu=-4$ (lower-left), and $\lg
\mu=-5$ (lower-right).} \label{fig:particles3_2}
\end{figure}

The most interesting pattern that is observed in the final
particle distribution has been obtained for the particles
initially distributed near the 3:2 resonance.
Fig.~\ref{fig:particles3_2} presents the final distributions in
four models with various values of the mass parameter $\mu$. One
may see that at $\lg\mu=-2$ the resulting pattern is strongly
blurred, whereas at $\lg\mu=-3$ it clearly appears as a
quasi-triangular one. On decreasing $\mu$, the regions of
increased particle density expand in azimuth, while the sizes of
cavities decrease. Therefore, in what follows we concentrate on
the possibility of actual astronomical observations of such a
pattern when $\lg\mu=-3$.

\subsection*{Calculation of theoretical images}

Sizes of particles in planetesimal disks vary from tens of
micrometers to hundreds of kilometers \citep{2008ARA&A..46..339W,
2010RAA....10..383K}. The size distribution is controlled by the
process of destructive collisions of planetesimals; and the fine
dust ($<10\mu$m) is removed from the disk by radiation pressure
\citep{2000A&A...362.1127K}. The collisional dynamics of
planetesimals was studied in a number of papers
\citep{2003ApJ...598..626D, 2006A&A...455..509K,
2007ApJ...658..569W, 2007A&A...472..169T, 2008ApJ...673.1123L,
2012ApJ...754...74G}. In particular, it was shown that the
collisional evolution of particles leads to their size
distribution in the form $dN=D^{2-3q}dD$ (where $q=5/3$~--~$2$);
and the mass distribution is $dN=m^{-q}dm$. Usually, in modeling,
$q=11/6$ is chosen:
\begin{equation}
dN=D^{-3.5}dD . \label{eq:ND}
\end{equation}
This $q$ value is provided in the infinite collisional cascade
model \citep{1969JGR....74.2531D}.

Then the mass fraction of particles of size $D$ depends on $D$ as
follows: $dm\propto D^{-0.5}dD$. Integrating this relation, one
finds that for the objects ranging in size from $100$~km to $1$~km
the total contribution to the disk mass is $\sim 90\%$. However,
thermal radiation in the submillimeter range, which can be
detected by the ALMA (Atacama Large Millimeter Array) radio
interferometer, is emitted by 0.1--10~mm particles
\citep{2022arXiv220203053M}.

\citet{2005Sci...307...68G, 2007A&A...472..169T} give estimates of
mass of the dust ($D\le 1$~cm) component in the typical debris
disk: $M_\mathrm{ dust}=0.001$--$0.1M_{\oplus}$. In our
calculations for the emission of debris disk particles located in
the resonance region, the fine dust mass $M_\mathrm{dust}$ acts as
a parameter of the problem. Three types of particles (with sizes
$100$~$\mu$m, $1$~mm, and $1$~cm) are considered. Mass
$M_\mathrm{dust}$ is distributed between them according to
Eq.~(\ref{eq:ND}). The mass of one particle of each type in our
calculations is determined as follows:
$m_\mathrm{D}=M_\mathrm{dust}/N_\mathrm{tot}\cdot f_\mathrm{D}$,
where $N_\mathrm{tot}$ is the total number of disk particles at
time $10^5P$, and $f_\mathrm{D}$ is the relative fraction of
particles of size $D$.

In the  calculations of particle dynamics, we consider the planar
(two-dimensional) problem; accordingly, in the radiation
calculations, each particle is assigned with a random value of the
coordinate $z$ in the range $[-0.005~$au$:0.005~$au$]$. Our
calculations show that variations in the choice of the range in
$z$ do not significantly affect the general character of the
emission of the structured disk.

The entire domain of computation is divided into cells in the
spherical coordinate system, namely,
$R\times\theta\times\phi=200\times4\times90$, where $R\in
[0.1a:4a]$, $\theta\in[88^\circ .85:91^\circ .15]$, $\phi\in
[0:2\pi)$. In each cell, the number of computed particles is
summed ($N_i$), then the concentration of particles of each type
in each cell is calculated:

$$ \rho_\mathrm{d} = \frac{m_\mathrm{D}\cdot N_i}{R^2
\sin(\theta) dR d\theta d\phi} . $$

In our three-dimensional computations of radiative transfer we
used the RADMC-3D
code\footnote{https://www.ita.uni-heidelberg.de/~dullemond/software/radmc-3d/}
\citep{2012ascl.soft02015D}. The number of photons in the
computations of the direct radiation and scattering was set to
$10^9$. The dust opacity for magnesium--iron silicates
\citep{1995A&A...300..503D} was calculated in the Mie theory
\citep{1908AnP...330..377M}, using the code by
\citep{1998asls.book.....B}; the RADMC-3D package includes this
code.

We used the calculated radiation fluxes for construction of model
images that can be potentially built using the ALMA radio
interferometer. The simulation was performed using the CASA 6.5
simulator\footnote{https://casa.nrao.edu/}
\citep{2012ASPC..461..849P} for the reference nominal (arbitrary)
position ($\alpha=04$h$33$m, $\delta=+22^\circ 53'$, J2000), by
analogy with the work by \cite{2015A&A...579A.110R}. Thermal noise
was added using the {\it tsys-atm} option of the CASA package,
with the water vapour deposited at $PWV=0.6$. It is assumed that
the observations are performed at the wavelength $870 \mu$m. The
bandwidth for observations in the continuum was set equal to $8$
GHz, the exposure time to 5~hr. The antenna configuration
corresponds to the largest compact configuration (\#20 of the
available CASA configurations).

\subsection*{Image analysis}

\begin{figure}
\centering \includegraphics[width=0.8\textwidth]{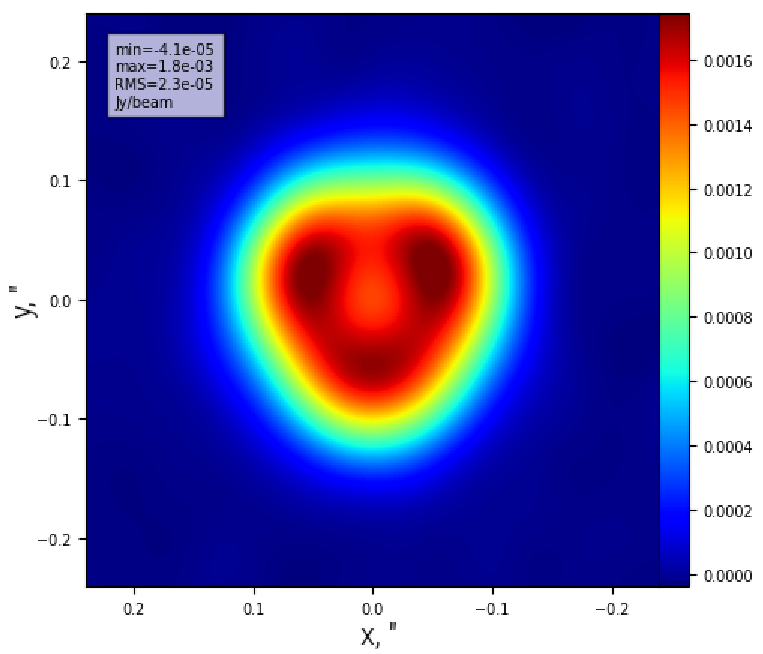}
\caption{The debris disk image at wavelength $870$~$\mu m$ in the
model with $a=1$~AU, $d=10$~pc, and
$M_\mathrm{dust}=0.001M_\oplus$. The color scale is given in
Jy/beam; its maximum and minimum, as well as the signal-to-noise
ratio, are indicated in the upper-left corner. The axial
coordinates are given in arcseconds.} \label{fig:1au}
\end{figure}

In our calculations of images of debris disk patterns, the input
parameters were as follows: mass $M_\mathrm{dust}$ of fine ($\le
1$~cm) dust, semimajor axis $a$ of the planet's orbit, and the
distance $d$ to the object. Fig.~\ref{fig:1au} presents a model
with $M_\mathrm{dust}=0.001M_\oplus$, $a=1$~AU, and $d=10$~pc. The
image obtained in the simulation looks blurred, but the bright
zone has a definite quasi-triangular shape.

\begin{figure}
\centering \includegraphics[width=0.8\textwidth]{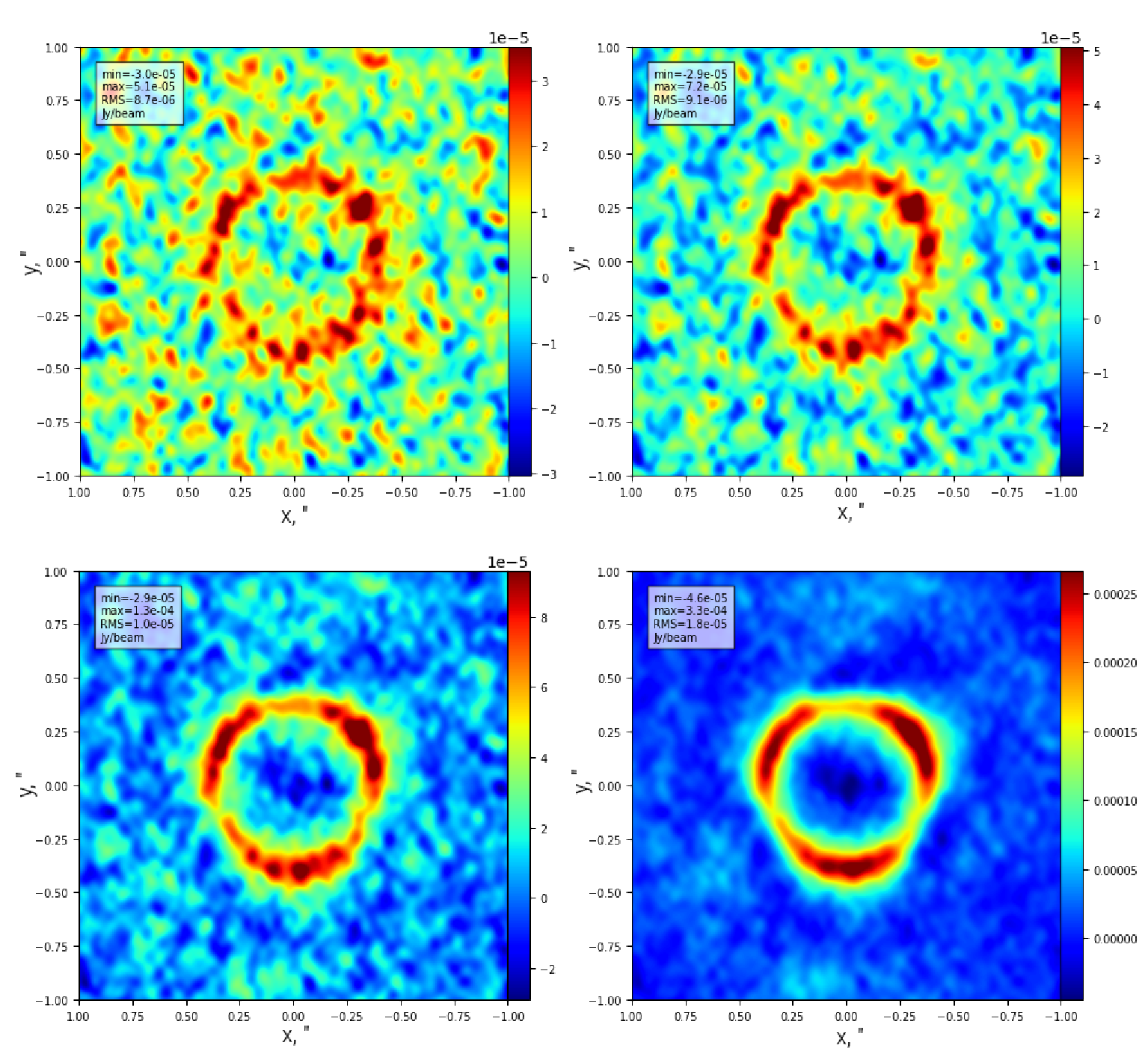}
\caption{Same as Fig.~\ref{fig:1au}, but in the models with
$a=5$~au and $d=10$~pc; $M_\mathrm{dust}=0.0003M_\oplus$
(upper-left), $M_\mathrm{dust}=0.005M_\oplus$ (upper-right),
$M_\mathrm{dust} =0.001M_\oplus$ (lower-left), and
$M_\mathrm{dust}=0.003M_\oplus$ (lower-right).} \label{fig:5au}
\end{figure}

Consequent series of images was obtained in the model with
$a=5$~AU and $d=10$~pc, while mass $M_\mathrm{dust}$ is varied
(Fig.~\ref{fig:5au}). One may see that, in the case of small
$M_\mathrm{dust} \le 0.0005M_\oplus$, the pattern is resolved
against the background noise, but it looks ragged; essentially, it
forms an annular density perturbation. On increasing
$M_\mathrm{dust}$, three bright spots emerge in the image (they
correspond to areas of increased density near the 3:2 resonance),
and the disk pattern acquires a characteristic quasi-triangular
shape.

\begin{figure}
\centering \includegraphics[width=0.8\textwidth]{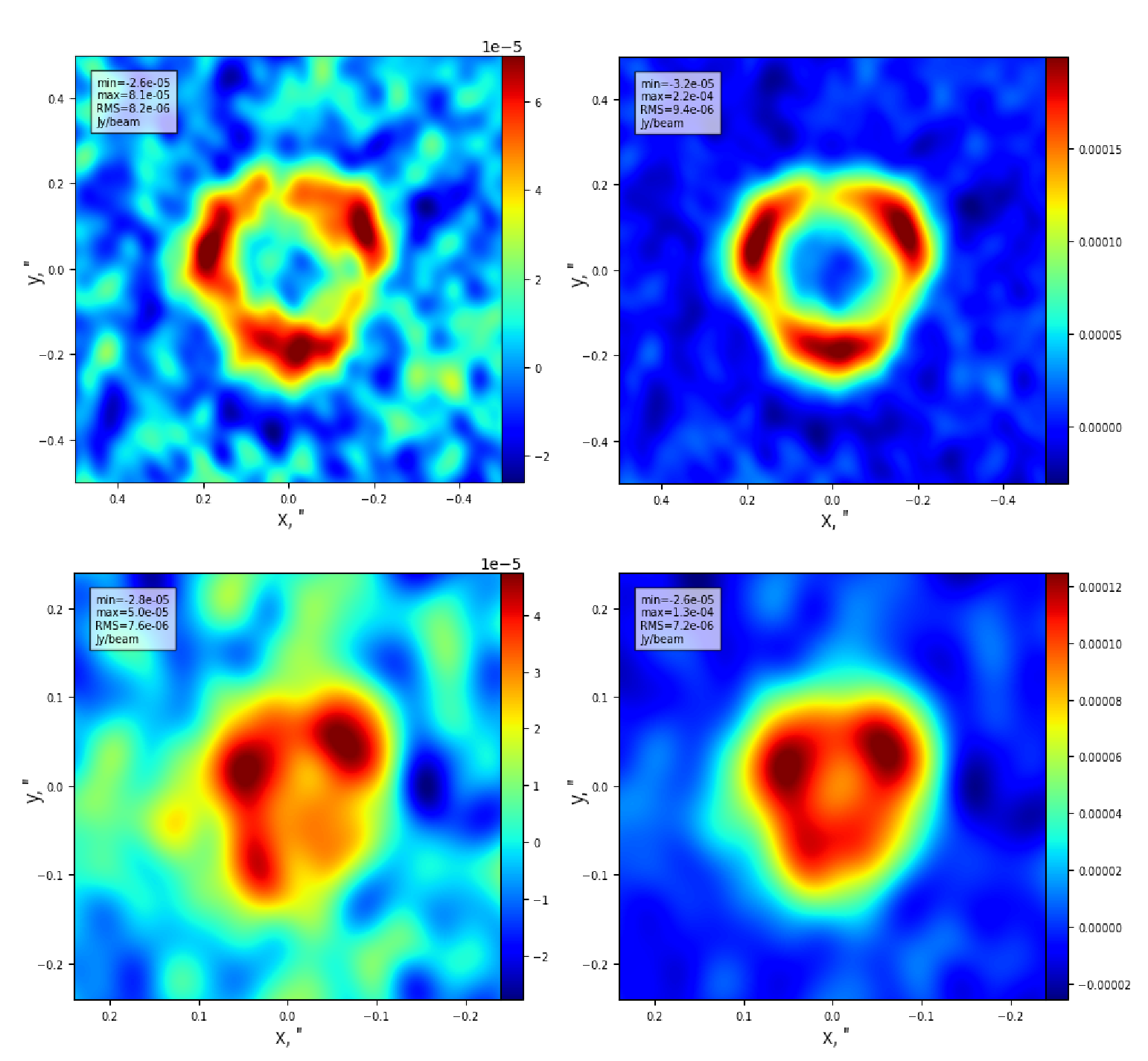}
\caption{Same as in Fig.~\ref{fig:1au}, but in the models with
$a=5$~au. The left-hand panels correspond to the models with
$M_\mathrm{dust}=0.001M_\oplus$, and the right-hand ones to those
with $M_\mathrm{dust}=0.003M_\oplus$. The observer--object
distance $d=20$~pc (upper panels) and $d=50$~pc (lower panels).}
\label{fig:bigD}
\end{figure}

Replacing the object to a greater distance $d$ from the observer
reduces the image resolution; however, even at $d=20$~pc, the
three bright spots corresponding to the disk matter clumps are
still quite distinguishable if $M_\mathrm{dust} \ge 0.001M_\oplus$
(Fig.~\ref{fig:bigD}, upper panels). When the object is replaced
to $50$~pc, the quasi-triangular pattern is distinguishable at
$M_\mathrm{dust} = 0.003M_\oplus$ (Fig.~\ref{fig:bigD}, lower
panels) and would be more clear-cut at larger $M_\mathrm{dust}$.

\section*{Conclusions}

Our calculations showed that the disk particles, initially placed
in zones of low-order mean-motion resonances with the planet,
eventually concentrate into noticeable azimuthal patterns formed
by thickening and cavities in the disk. In all considered models,
the low-density cavities emerge close to the star and the planet.
The structuring process usually proceeds on the time scale
$\sim$100, in units of the planet orbital periods. The number of
particles that retain their locations near the resonances
increases with decreasing the mass parameter $\mu$.

However, in all models, a removal of a significant fraction of
particles from the disk is substantial due to their accretion onto
the star and the planet, as well as due to transition to highly
elongated and hyperbolic orbits. Particles can be transferred to
hyperbolic orbits from the zones of all considered resonances, if
$\lg\mu > -5$.

In the considered models, the azimuthal pattern associated with
the 3:2 resonance manifests itself most clearly (if one sets
$\lg\mu=-3$, corresponding to the mass ratio in the
``Sun--Jupiter'' system). Observability of such a pattern is real
solely for the stars in the nearest Galactic neighborhood of the
Sun; if $a>5$~AU and $d>50$~pc, the observability requires
presence of fine dust of significant mass in the disk. However, it
should be noted that the value of the mass $M_\mathrm{dust}$
accepted by us in the calculations does not exceed $3\%$ of the
maximum possible mass of fine dust in the debris disk.

Our simulations also revealed an azimuthal pattern associated with
another considered first-order resonance, namely, the 2:1
resonance. Visual examples of actually observed azimuthal
patterns, similar to those produced by this resonance, are
provided by the protoplanetary circumstellar disks of the stars
HD~169142 \citep{2017A&A...600A..72F}, HD~97804
\citep{2017A&A...597A..32V}, and AA~Tau
\citep{2017ApJ...840...23L}; see also Fig.~1 in the review
\citep{2018ASPC..517..137A}. Their possible correspondence with
theoretical resonance models deserves separate analysis.

If one succeeds in the future to identify the presented above
theoretical azimuthal patterns in the observed images of
circumstellar disks, then, by modeling the image in each case, it
would be possible to estimate the mass parameter $\mu$ and the
orbital semimajor axis $a$ of the planet that retains the
particles in the resonance zone. For example, if a
quasi-triangular bright pattern is detected in the image, the mass
parameter $\mu$ can be estimated, based on results of our
constructing the model images, as $\gtrsim 10^{-3}$, and the
semimajor axis of the planet's orbit
$a\approx\Big(\frac{3}{2}\Big)^\frac{2}{3}a_t\approx 1.3a_t$,
where $a_t$ corresponds to radial location of the centers of the
bright patterns.

\section*{Acknowledgments}
The computations were carried out using the resources of the Joint
Supercomputer Center of the Russian Academy of Sciences --- Branch
of Federal State Institution ``Scientific Research Institute for
System Analysis of the Russian Academy of
Sciences''\footnote{https://www.jscc.ru/}
\citep{2019LG..40..1835}. The study was supported by the Russian
Foundation for Basic Research, project No.~22-22-00046.

\newpage

\bibliographystyle{plainnat}
\bibliography{biblio}

\newpage

\end{document}